\documentclass[12pt]{article}
\usepackage{amsfonts,amsmath,amssymb}

\def\bq{ \begin{equation} }
\def\eq{ \end{equation} }
\def\ben{ \begin{eqnarray} }
\def\en{ \end{eqnarray} }

\def\frac#1#2{{#1\over #2}}

\def\on#1#2{\mathop{\vbox{\ialign{##\crcr\noalign{\kern2pt}
$\scriptstyle{#2}$\crcr\noalign{\kern2pt\nointerlineskip}
\kern-2pt$\hfil\displaystyle{#1}\hfil$\crcr}}}\limits}

\begin{document}

\baselineskip=15pt
\vspace{1cm} \noindent {\LARGE \textbf{Bi-Hamiltonian ODEs with matrix variables}} \vskip1cm \hfill
\begin{minipage}{13.5cm}
\baselineskip=15pt
{\bf A Odesskii  ${}^{1}$,
    V Rubtsov ${}^{2}$ and
   V Sokolov ${}^{3}$} \\ [2ex]
{\footnotesize
${}^1$  Brock University, St. Catharines (Canada)
\\
${}^2$ Institute for Theoretical and Experimental Physics, Moscow (Russia) and LAREMA, UMR 6093 du CNRS, Angers University (France)
\\
${}^{3}$ Landau Institute for Theoretical Physics, Moscow (Russia) }\\
\vskip1cm{\bf Abstract.} We consider a special class of linear and quadratic Poisson brackets related to
ODE systems with matrix variables. We investigate general properties of such brackets, present an example of a compatible pair
of quadratic and linear brackets and found the corresponding hierarchy of integrable models, which generalizes the two-component Manakov's matrix system in the case of arbitrary number of matrices.

\end{minipage}

\vskip0.8cm
\noindent{
MSC numbers: 17B80, 17B63, 32L81, 14H70 }
\vglue1cm \textbf{Address}:
Landau Institute for Theoretical Physics, Kosygina 2, 119334, Moscow, Russia

\textbf{E-mail}:
sokolov@itp.ac.ru, \, aodesski@brocku.ca, \, volodya@univ-angers.fr \newpage

\section{Introduction}
\medskip

We consider ODE systems of the form
\begin{equation}\label{geneq}
\frac{d x_{\alpha}}{d t}=F_{\alpha}({\bf x}), \quad {\bf x}=(x_1,...,x_N),
\end{equation}
where $x_i$ are $m\times m$-matrices and $F_{\alpha}$ are (non-commutative) polynomials.
There exist systems (\ref{geneq}) integrable for any $m$. For example, the system
\begin{equation}\label{man}
u_{t}=u^2 \, v-v \, u^2, \qquad v_{t}=0
\end{equation}
is integrable by the Inverse Scattering Method for any size $m$ of matrices $u$ and $v$.
If $u$ is a matrix such that $u^T=-u$, and $v$ is a constant diagonal matrix, then
(\ref{man}) is equivalent to the $m$-dimensional Euler top. The
integrability of this model was established by S.V. Manakov
in 1976 (\cite{man}).

In this paper we construct an integrable generalization of system (\ref{man}) to the case of arbitrary $N$ using the bi-Hamiltonian approach \cite{magri}. This approach is based on the notion of a pair of compatible Poisson brackets. Two Poisson brackets $\{\cdot, \cdot \}_1$ and $\{\cdot, \cdot
\}_2$ are said to be compatible if
\begin{equation}\label{pencil}
\{\cdot, \cdot \}_{\lambda}=\{\cdot, \cdot \}_1 +\lambda \{\cdot,
\cdot \}_2
\end{equation}
is a Poisson bracket for any constant $\lambda$.

If the bracket (\ref{pencil}) is degenerate, then a hierarchy of integrable Hamiltonian ODE systems can be constructed via the following

{\bf Theorem 1} (\cite{magri1, zakhar}). Let
$$
C(\lambda)=C_0+\lambda C_1+\lambda^2 C_2+\cdots, \qquad \bar C(\lambda)=\bar C_0+\lambda \bar C_1+\lambda^2 \bar C_2+\cdots,
$$
be Taylor expansion of any two Casimir functions for the bracket $\{\cdot, \cdot
\}_{\lambda}$. Then the coefficients $C_i, \bar C_j$ are pairwise commuting with
respect to both brackets $\{\cdot, \cdot \}_1$ and $\{\cdot, \cdot
\}_2.$

In the opposite case when, say, the bracket $\{\cdot, \cdot
\}_1$ is nondegenerate, there is another way to construct an integrable hierarchy. The ratio $R=\Pi_2 \Pi_1^{-1},$ where $\Pi_i$ is the Poisson tensor for $\{\cdot, \cdot\}_i$
defines a so-called recursion operator, whose spectrum provides the set of functions in involution with respect to both brackets. In this case the formula $\Pi_k=R^k \Pi_1$ gives us an infinite sequence of pairwise compatible Poisson brackets.

For an important class of Poisson brackets related to systems (\ref{geneq}) the corresponding Hamiltonian operator can be expressed in terms of left and right multiplication
operators given by polynomials in $x_1,...,x_N$ \cite{MikSok}.  Such brackets possess the following two properties:
\begin{itemize}
\item  they are $GL_m$-adjoint invariant;
\item the bracket between traces of any two matrix polynomials $P_i(x_1,...,x_N), \quad i=1,2$ is a trace of some other matrix polynomial $P_3$.
\end{itemize}

Such brackets we shall call {\it non-abelian Poisson brackets}

In this paper we consider compatible pairs of non-abelian Poisson brackets, where the bracket $\{\cdot, \cdot \}_1$ is linear and  $\{\cdot, \cdot
\}_2$ is  quadratic.

\section{ Non-abelian Poisson brackets}

We consider Poisson brackets of the following form:
\begin{equation}\label{Poissonlin}
\{x^j_{i,\alpha},x^{j^{\prime}}_{i^{\prime},\beta}\}=
b_{\alpha,\beta}^{\gamma}x_{i,\gamma}^{j^{\prime}}\delta^j_{i^{\prime}}-
b_{\beta,\alpha}^{\gamma}x_{i^{\prime},\gamma}^j\delta^{j^{\prime}}_i,
\end{equation}
and
\begin{equation}\label{Poisson}
\{x^j_{i,\alpha},x^{j^{\prime}}_{i^{\prime},\beta}\}=
r^{\gamma\epsilon}_{\alpha\beta}x^{j^{\prime}}_{i,\gamma}x^j_{i^{\prime},\epsilon}+
a^{\gamma\epsilon}_{\alpha\beta}x^k_{i,\gamma}x^{j^{\prime}}_{k,\epsilon}\delta^j_{i^{\prime}}-
a^{\gamma\epsilon}_{\beta\alpha}x^k_{i^{\prime},\gamma}x^{j}_{k,\epsilon}\delta^{j^{\prime}}_i,
\end{equation}
where $x^j_{i,\alpha}$ are entries of the matrix $x_{\alpha}$ and $\delta^{j}_i$ is the Kronecker delta.  The summation with respect to repeated indexes is assumed.
Here and in the sequel we use Latin indexes for the entries of matrices. They vary from 1 to $m$. The Greek indexes varying from 1 to $N$  are used for numbering of matrices.

{\bf Theorem 2.}  Brackets of the form (\ref{Poissonlin}) and (\ref{Poisson}) are both invariant with respect to $GL_m$-action $x_{\alpha}\to u x_{\alpha} u^{-1},$ where $u\in GL_m.$ Moreover, these brackets satisfy the following property: the bracket between traces of any two matrix polynomials is a trace of a matrix polynomial. Any linear (respectively quadratic) Poisson bracket satisfying these two properties has the form (\ref{Poissonlin}) (respectively (\ref{Poisson})).

There are a lot of publications devoted to quadratic Poisson brackets that appeared in the classical version of Inverse
Scattering Method \cite{reysem}. However, these brackets do not satisfy the properties of Theorem 2.

{\bf Theorem 3.} 1) Formula (\ref{Poissonlin}) defines a Poisson
bracket iff
\begin{equation}\label{r0}
b^{\mu}_{\alpha \beta} b^{\sigma}_{\mu \gamma}=b^{\sigma}_{\alpha \mu} b^{\mu}_{\beta \gamma};
\end{equation}

2) Formula (\ref{Poisson}) define a Poisson bracket iff the following relations hold:
\begin{equation}\label{r1}
r^{\sigma \epsilon}_{\alpha\beta}=-r^{\epsilon\sigma}_{\beta\alpha},
\end{equation}
\begin{equation}\label{r2}
r^{\lambda\sigma}_{\alpha\beta}
r^{\mu\nu}_{\sigma\tau}+r^{\mu\sigma}_{\beta\tau} r^{\nu\lambda}_{\sigma\alpha}+r^{\nu\sigma}_{\tau\alpha} r^{\lambda\mu}_{\sigma\beta}=0,
\end{equation}
\begin{equation}\label{r3}
a^{\sigma\lambda}_{\alpha\beta} a^{\mu\nu}_{\tau\sigma}=a^{\mu\sigma}_{\tau\alpha} a^{\nu\lambda}_{\sigma\beta},
\end{equation}
\begin{equation}\label{r4}
a^{\sigma\lambda}_{\alpha\beta} a^{\mu\nu}_{\sigma\tau}=a^{\mu\sigma}_{\alpha\beta} r^{\lambda\nu}_{\tau\sigma}+a^{\mu\nu}_{\alpha\sigma}
r^{\sigma\lambda}_{\beta\tau}.
\end{equation}
and
\begin{equation}\label{r5}
a^{\lambda\sigma}_{\alpha\beta} a^{\mu\nu}_{\tau\sigma}=a^{\sigma\nu}_{\alpha\beta} r^{\lambda\mu}_{\sigma\tau}+a^{\mu\nu}_{\sigma\beta}
r^{\sigma\lambda}_{\tau\alpha}.
\end{equation}

Formula (\ref{r0}) mean that  $b^{\sigma}_{\alpha \beta}$
are structure constants of an associative algebra. Similar Poisson brackets were considered in (\cite{OdSok1},\cite{PVdW}). Let us consider the quadratic Poisson brackets (\ref{Poisson}).

Under change of basis $x_\alpha \to g^\beta_\alpha x_\beta$ the constants are
transformed in a standard way:
\begin{equation}\label{basis}
r_{\alpha\beta}^{\gamma \sigma} \to g^{\lambda}_\alpha g^{\mu}_\beta h_{\nu}^\gamma h_{\epsilon}^\sigma
\,r_{\lambda \mu}^{\nu \epsilon}, \qquad
a_{\alpha\beta}^{\gamma\sigma} \to g^{\lambda}_\alpha g^{\mu}_\beta h_{\nu}^\gamma h_{\epsilon}^\sigma
\,a_{\lambda \mu}^{\nu \epsilon},
\end{equation}
Here $g_\alpha^\beta h_\beta^\gamma=\delta_\alpha^\gamma$.

The system  of identities (\ref{r1})-(\ref{r5})  admits the following
discrete involution:
\begin{equation}\label{tran2}  r^{\gamma\sigma}_{\alpha\beta}\to r^{\gamma\sigma}_{\beta\alpha},\qquad
a^{\gamma\sigma}_{\alpha\beta}\to a^{\sigma\gamma}_{\beta\alpha}.\end{equation}
This involution corresponds to the matrix transposition $x_\alpha\to x_\alpha^T.$
Two brackets related by (\ref{basis}),(\ref{tran2}) are called {\it equivalent}.

There exists one more discrete involution:
\begin{equation}\label{tran1} r^{\gamma\sigma}_{\alpha\beta}\to r_{\gamma\sigma}^{\alpha\beta},\qquad
a^{\gamma\sigma}_{\alpha\beta}\to a_{\gamma\sigma}^{\alpha\beta}. \end{equation}
The brackets related by (\ref{tran1}) in
principle may have
different properties, for example different sets of Casimir functions.

Let $V$ be a linear space with a basis $v_\alpha,~\alpha=1,...,N$. Define linear
operators $r,~a$ on the space $V\otimes V$ by $$rv_\alpha\otimes
v_\beta=r^{\sigma \epsilon}_{\alpha \beta}v_\sigma\otimes v_\epsilon,\quad av_\alpha \otimes v_\beta=a^{\sigma \epsilon}_{\alpha \beta}v_\sigma\otimes
v_\epsilon.$$  Then the identities (\ref{r1})-(\ref{r5}) can be rewritten in
the following form:
$$r^{12}=-r^{21},~~~r^{23}r^{12}+r^{31}r^{23}+r^{12}r^{31}=0,$$
$$a^{12}a^{31}=a^{31}a^{12},$$
$$\sigma^{23}a^{13}a^{12}=a^{12}r^{23}-r^{23}a^{12},$$
$$a^{32}a^{12}=r^{13}a^{12}-a^{32}r^{13}.$$
Here all operators act in $V\otimes V\otimes V$, by $\sigma^{ij}$ we
mean the transposition of $i$-th and $j$-th component of the tensor
product and $a^{ij},~r^{ij}$ mean operators $a,~r$ acting in the
product of the $i$-th and $j$-th components.

The involution (\ref{tran1}) corresponds to  $a\to a^*, \, r\to r^*$,
where $a^*,~r^*$
act in dual space $V^*$; (\ref{tran2}) corresponds to $a\to \sigma a\sigma, \,
r\to \sigma r$ where $\sigma$ acts by the permutation of vector spaces
in the space $V\otimes V$. The equivalence transformation
(\ref{basis}) corresponds to $a \to GaG^{-1},\,
r \to GrG^{-1}$, where $G=g\otimes g$ and $g\in GL(V)$.

There is a subclass of brackets (\ref{Poisson}) that corresponds to
the tensor $a$ equals to $0$. Relations  (\ref{r1}), (\ref{r2}) mean
that $r$ is a constant solution of the associative Yang-Baxter
equation (\cite{Agui},\cite{Sch}). Such tensors $r$ can be
constructed in the following algebraic way.

An {\it anti-Frobenius algebra} is an associative algebra  $\cal A$
(not necessarily with unity) with non-degenerate
anti-symmetric bilinear form $(~,~)$ satisfying the following relation
\begin{equation}\label{af}
(x,yz)+(y,zx)+(z,xy)=0
\end{equation}
for all $x,y,z\in \cal A$. In other words the form $(~,~)$ defines a cyclic 1-cocycle on $\cal A$.

{\bf Theorem 4.} There exists one-to-one correspondence between
solutions of (\ref{r1}), (\ref{r2})  up to equivalence and exact
representations of anti-Frobenius algebras up to isomorphism.

{\bf Proof.} Tensor $r$ can be written as
$r^{ij}_{kl}=\sum_{\alpha,\beta=1}^pg^{\alpha\beta}y^i_{k,\alpha}y^j_{l,\beta},$
where $g^{\alpha\beta}=-g^{\beta\alpha}$, the matrix $G=(g^{\alpha\beta})$
is non-degenerate and $p$ is the smallest possible.
Substituting this representation into (\ref{r1}), (\ref{r2}), we obtain that there
exists a tensor $\phi^{\gamma}_{\alpha\beta}$
such that
$y^i_{k,\alpha}y^k_{j,\beta}=\phi^{\gamma}_{\alpha\beta}y^i_{j,\gamma}$. Let
 $\cal A$ be the
associative algebra with the basis $y_1,...,y_p$ and the product
$y_{\alpha}y_{\beta}=\phi^{\gamma}_{\alpha\beta}y_{\gamma}$.
Define the anti-symmetric bilinear form by $(y_{\alpha},y_{\beta})=g_{\alpha\beta},$ where
$(g_{\alpha\beta})=G^{-1}$.
Then (\ref{r1}), (\ref{r2}) is equivalent to anti-Frobenius property (\ref{af}).

{\bf Example 1} \,(cf. \cite{elash}). Let $\cal A$ be associative algebra of $N\times N$-matrices with zero $N$-th row, $l$ be
generic element of ${\cal A}^{*}$. Then $(x,y)=l([x,y])$ is a non-degenerate
anti-symmetric bilinear form satisfying (\ref{af}). Let $(x,y)={\rm trace}([x,y]\, k^T),$ where $k\in {\cal A}$. Then we can put $k_{ij}=0, i\ne j$, $k_{ii}=\mu_i,$ where $i,j=1,...,N-1,$ and $k_{iN}=1,\, i=1,...,N-1$. The corresponding bracket (\ref{Poisson}) is given by the following tensor $r$:
\begin{equation}\label{elash}
r^{ii}_{Ni}=-r^{ii}_{iN}=1, \qquad r^{ij}_{ij}=r^{ji}_{ij}=r^{ii}_{ji}=-r^{ii}_{ij}=\frac{1}{\mu_i-\mu_j},
\quad i\ne j,\quad i,j=1,...,N-1.
\end{equation}
The remaining elements of the tensor $r$ and all elements of the tensor $a$ supposed to be zero. Notice that this tensor is anti-symmetric with respect to involution (\ref{tran2}). The bracket (\ref{elash}) is equivalent to
\begin{equation}\label{ex4}
  r^{\alpha \beta}_{\alpha \beta}=r^{\beta \alpha}_{\alpha \beta}=r^{\alpha\alpha}_{\beta\alpha}=-r^{\alpha\alpha}_{\alpha\beta}=\frac{1}{\lambda_\alpha-\lambda_\beta},
\quad \alpha\ne \beta,\quad \alpha,\beta=1,\ldots,N.
\end{equation}
Here $\lambda_1,\ldots,\lambda_N$ are arbitrary pairwise distinct parameters. For $m=1$ we have the following scalar Poisson bracket
$$
\{x_\alpha,x_\beta\}=\frac{(x_\alpha-x_\beta)^2}{\lambda_\beta-\lambda_\alpha}, \quad \alpha\ne \beta,\quad \alpha,\beta=1,...,N.
$$

 If $N$ is even, then the Poisson structure (\ref{ex4}) is non-degenerate, i.e. the rank of the Poisson tensor $\Pi$ is equal to $
N m^2$. In the odd case $rank \, \Pi =(N-1) m^2$.

There is the following known way (the so-called argument shift method) for constructing a linear Poisson bracket compatible
with a quadratic one.
A vector $\mu=(\mu_1,...,\mu_m)$ is said to be {\it
admissible} if for any $\alpha,\beta$
$$
(a^{\sigma \epsilon}_{\alpha \beta}  - a^{\epsilon\sigma}_{\beta\alpha} + r^{\sigma \epsilon}_{\alpha \beta}) \mu_\sigma \mu_\epsilon=0.
$$
For any admissible vector the argument shift $x_\alpha \to x_\alpha+\mu_\alpha \,
{\rm Id}$ yields a linear Poisson bracket with
$$
b^{\sigma}_{\alpha \beta}=(a_{\alpha \beta}^{\epsilon\sigma}+a_{\alpha \beta}^{\sigma \epsilon}+r_{\alpha \beta}^{\sigma \epsilon})\mu_\epsilon,
$$
compatible with the quadratic one. For Example 1 any admissible vector is proportional to $(1,1,...,1)$ and the corresponding linear bracket is trivial.

{\bf Example 2.} Applying the involution (\ref{tran1}) to
(\ref{ex4}), we get one more example with zero tensor $a$:
\begin{equation}\label{manbr}
  r_{\alpha \beta}^{\alpha \beta}=r_{\beta \alpha}^{\alpha \beta}=r_{\alpha\alpha}^{\beta\alpha}=-r_{\alpha\alpha}^{\alpha\beta}=\frac{1}{\lambda_\alpha-\lambda_\beta},
\quad \alpha\ne \beta,\quad \alpha,\beta=1,\ldots,N.
\end{equation}
It is easy to verify that in this case any vector $(\mu_1,...,\mu_N)$ is admissible. All entries of the matrix $\sum_1^N x_\alpha$ are Casimir functions for both quadratic Poisson bracket $\{\cdot,\cdot \}_2$
of Example 2 and for the corresponding linear bracket $\{\cdot,\cdot \}_1.$ Thus we can fix
$$
\sum_1^N x_\alpha=C,
$$
where $C$ is a constant matrix.
Hamiltonians of the hierarchy commuting with respect to  both $\{\cdot,\cdot \}_2$ and  $\{\cdot,\cdot \}_1$  are given by
$$
{\rm tr}\, x_\alpha^k, \qquad {\rm tr}\,x_{\alpha}^{k} \sum_{\beta\ne \alpha} \frac{x_{\beta}}{\lambda_{\alpha}-\lambda_{\beta}}, \qquad k=1,2,....
$$
The Casimir functions of $\{\cdot,\cdot \}_1$ belong to this set, which gives a constructive way to find the whole hierarchy.

The dynamical system corresponding to the simplest Hamiltonian ${\rm tr}\, x_N$ and the Poisson structure
$\{\cdot,\cdot \}_2$ has the form
$$
\frac{d x_\alpha}{d t}=\frac{x_N x_\alpha-x_\alpha x_N}{\lambda_N-\lambda_\alpha},
\quad \alpha=1,\ldots,N-1.
$$

The linear Casimir functions for  $\{\cdot,\cdot \}_1$ are ${\rm tr}\, x_\alpha,$ where
$\alpha=1,\ldots,N.$ There exists the following quadratic Casimir function:
$$
H=\frac{1}{2} \sum_{\alpha=1}^{N} \frac{1}{\mu_\alpha} {\rm tr}\, x_\alpha^2.
$$
The  non-abelian system corresponding to this Hamiltonian and the Poisson bracket
$\{\cdot,\cdot \}_2$  is given by
\begin{equation}\label{sys}
\frac{d x_\alpha}{d t}=\sum_{\beta\ne \alpha} \frac{x_\alpha x_\beta^2-x_\beta^2 x_\alpha}{(\lambda_\alpha-\lambda_\beta) \mu_\beta}+\sum_{\beta\ne \alpha} \frac{x_\beta x_\alpha^2-x_\alpha^2 x_\beta}{(\lambda_\alpha-\lambda_\beta) \mu_\alpha}.
\end{equation}
The system (\ref{sys}) can be written in the following bi-Hamiltonian form:
$$
\frac{d {\bf x}}{d t}=\{{\bf x}, {\rm grad\,(tr}\,H) \}_2=\{{\bf x}, {\rm grad\,(tr}\,K) \}_1,
$$
where
$$
K=\frac{1}{3} \sum_{\alpha=1}^{N} \frac{1}{\mu_\alpha^2} {\rm tr}\, x_\alpha^3.
$$
If $N=2$, then system (\ref{sys}) is equivalent to (\ref{man}).

\section{ Conclusions and outlooks}

We have proposed some examples of linear and quadratic Poisson
brackets naturally related to some matrix ODE. Similar Poisson,
symplectic and many other interesting algebraic structures were
appeared recently in the framework of the M.Kontsevich's approach to
the Non-Commutative Symplectic Geometry (\cite{Konts}). We mention
here the works of P.Etingof, V. Ginzburg, W. Crawley-Boevey
(\cite{ECG},\cite{CB}), L. Le Bryun (\cite{LB}), M. Van den Bergh
(\cite{VdB}) and many others on Calogero-Moser spaces, symplectic
and Poisson geometry of Quiver Path algebras, necklace and double
Poisson structures, Leibniz-Loday algebras, Rota-Baxter algebras
etc.

A forthcoming paper \cite{OdRubSok} establishes the place of our
non-abelian Poisson structures in the realm of these non-commutative
algebro-geometric notions.We shall discuss the relations of our
quadratic Poisson structures to the Poisson geometry of the affine
variety associated with the representation spaces modulo adjoint
$GL-$action, describe classification results for quadratic brackets
for a free associative algebra case and the link with the
``quadratic'' double Poisson structures. We shall describe their
symplectic foliations and Casimirs as well as the correspondent
integrable systems.

Another interesting and intriguing problem relates to the
quantization of the non-abelian brackets, their relations to
different versions of dynamical Yang-Baxter equations and their
generalizations, the quantization of the non-abelian integrable
equations. All that is beyond the scope of the papers.  We hope to
return to this subject elsewhere.

\vskip.3cm \noindent {\bf Acknowledgments.} The authors are grateful
to A. Alekseev, J. Avan, E.B. Vinberg, M. Kontsevich, A. Polishchuk
and M. Semenov-{T}ian-{S}hansky for useful discussions. VS is
grateful to Angers, Brock and Dijon Universities for hospitality
while the paper was written. VS and VR are grateful to MATPYL
project "Non-commutative integrable systems" and the ANR ``DIADEMS''
project for a financial support of VS visits in Angers. They were
also partially supported by the RFBI grant 11-01-00341-a.


\newpage
\begin{thebibliography}{10}

\bibitem{magri}F.~Magri,
\newblock{A simple model of the integrable Hamiltonian equation},
\newblock{\em J. Math. Phys.}, {\bf 19}, (1978) 1156--1162.

\bibitem{magri1}F.~Magri, P.~Casati, G.~Falqui, M.~Pedroni,
\newblock {\em Eight lectures on Integrable Systems}, In: Integrability of
Nonlinear Systems (Y. Kosmann-Schwarzbach et al. eds.), Lecture
Notes in Physics {\bf 495} (2nd edition), 2004, pp.\ 209-250.

\bibitem{zakhar}   I. M.~Gelfand,   I.~Zakharevich,
\newblock{ Lenard schemes, and the local geometry of bi-Hamiltonian Toda and Lax structures}.
\newblock{\em Selecta Math.} (N.S.) {\bf 6}, (2000), no. 2, 131–183.


\bibitem{elash}  A. G.~Elashvili,
\newblock{Frobenius Lie algebras.} (Russian)
\newblock{\em Funktsional. Anal. i Prilozhen. 16 (1982), no. 4, 94–95. {English
translation: Functional Anal. Appl.}16 (1982), no. 4, 326--328
(1983).}

\bibitem{Agui} M.~Aguiar,
\newblock{On the associative analog of Lie bialgebras},
\newblock{\em J.Alg.}, {\bf 244} (2001), no. 2, 492-532.

\bibitem{reysem} A.G. Reyman and M.A. Semenov-{T}ian-{S}hansky,
\newblock{Compatible Poisson structures for Lax equations: an $r$-matrix approach},
\newblock{\em Phys. Lett. A }, {\bf 130}(8-9), 456--460, 1988.

\bibitem{MikSok}  A.V.~Mikhailov  and V.V.~Sokolov,
\newblock{Integrable ODEs on Associative Algebras}. {\em Comm. Math. Phys}. {\bf 211}, 231-251, 2000.

\bibitem{man} S.V.~Manakov, \newblock{ Note on the integration of Euler's
equations of the dynamics of an n-dimensional rigid body}, \newblock{\em Funct. Anal.
Appl.} {\bf 10}, no.{\bf 4} (1976), 93-94

\bibitem{OdSok1}  A.V.~Odesskii  and V.V.~Sokolov,
\newblock{Integrable matrix
 equations related to pairs of compatible associative algebras},
\newblock{\em  Journal Phys. A: Math. Gen.},  2006, {\bf 39}, 12447--12456.


\bibitem{Konts} M.~Kontsevich,
\newblock{Formal (non)commutative symplectic geometry.}
\newblock{\em The Gel'fand Mathematical Seminars, 1990--1992},
173--187, Birkh\"{a}user Boston, Boston, MA, 1993.

\bibitem{VdB} M.~Van den Bergh,
\newblock{Double Poisson algebras.}
\newblock{\em Trans. Amer. Math. Soc.} {\bf 360}, (2008), no. 11, 5711–5769.

\bibitem{LB} L.~Le Bruyn,
\newblock{\em Noncommutative geometry and Cayley-smooth orders.} Pure and
Applied Mathematics (Boca Raton), 290. Chapman and Hall/CRC, Boca
Raton, FL, 2008. lxiv+524 pp.

\bibitem{ECG} W.~Crawley-Boevey, P.~Etingof, P.~Ginzburg,
\newblock{Noncommutative geometry and quiver algebras.}
\newblock{\em Adv. Math.} 209 (2007), no. 1, 274–336.

\bibitem{CB} W.~Crawley-Boevey,
\newblock{Poisson structure on moduli spaces of representations.}
\newblock{\em J. Alg.}{\bf 325}(2011),205--215.

\bibitem{PVdW} A.~Pichereau and G.~Van den Weyer,
\newblock{Double Poisson cohomology of path algebras of quivers.}
\newblock{\em J. Algebra}, {\bf 319} (2008), no. 5, 2166--2208.


\bibitem{OdRubSok} A.V.~Odesskii, V.~Rubtsov and V. V.~Sokolov
{\newblock Non-abelian quadratic Poisson brackets: algebraic aspects},
\newblock{\em in preparation}

\bibitem{Sch} T.~Schedler,
\newblock{Trigonometric solutions of the associative Yang-Baxter equation},
\newblock{\em  Math. Res. Lett.}, {\bf 10 }, 2003, no. 2-3, 301-321.

\end{thebibliography}
\end{document}